\documentclass[sigconf,screen]{acmart}
\usepackage{colortbl}
\usepackage{multirow}
\usepackage[inkscapelatex=false]{svg}
\usepackage{tcolorbox}
\usepackage{xcolor}
\usepackage{xfp}
\usepackage{microtype}
\usepackage{enumitem}
\usepackage{subfiles}
\usepackage{url}
\definecolor{graya}{HTML}{e6e6e6}
\def\mybar#1{
   {\color{black}\rule{\fpeval{#1/\percentscale*\barwidth} cm}{\barheight}\color{graya}\rule{\fpeval{(\percentscale-#1)/\percentscale*\barwidth} cm}{\barheight}} 
}
\newcommand{\barwidth}{0.8} 
\newcommand{\barheight}{4pt} 
\newcommand{\percentscale}{265.7} 
\def\BibTeX{{\rm B\kern-.05em{\sc i\kern-.025em b}\kern-.08em
    T\kern-.1667em\lower.7ex\hbox{E}\kern-.125emX}}

\usepackage{balance}

\usepackage[font=small,skip=2pt]{caption}
\newcommand{\distance}{2pt}
\newcommand{\revised}[1]{\textcolor{black}{#1}}

\AtBeginDocument{%
  \providecommand\BibTeX{{%
    \normalfont B\kern-0.5em{\scshape i\kern-0.25em b}\kern-0.8em\TeX}}}

\copyrightyear{2023} 
\acmYear{2023} 
\setcopyright{acmlicensed}
\acmConference[ISSTA '23]{Proceedings of the 32nd ACM SIGSOFT International Symposium on Software Testing and Analysis}{July 17--21, 2023}{Seattle, WA, United States}
\acmBooktitle{Proceedings of the 32nd ACM SIGSOFT International Symposium on Software Testing and Analysis (ISSTA '23), July 17--21, 2023, Seattle, WA, United States}
\acmPrice{15.00}
\acmDOI{10.1145/3597926.3598056}
\acmISBN{979-8-4007-0221-1/23/07}
\acmSubmissionID{issta23main-p85-p}
\received{2023-02-16}
\received[accepted]{2023-05-03}

\begin{document}

\title{A Comprehensive Study on Quality Assurance Tools for Java}




\author{Han Liu}
\affiliation{%
  \institution{East China Normal University}
  \city{Shanghai}
  \country{China}
}
\authornote{These authors contributed equally to this work.}

\email{hanliu@stu.ecnu.edu.cn}

\author{Sen Chen}
\authornotemark[1]
\affiliation{%
  \institution{College of Intelligence and Computing, Tianjin University}
  \city{Tianjin}
  \country{China}
}
\email{senchen@tju.edu.cn}

\author{Ruitao Feng}
\affiliation{%
  \institution{University of New South Wales}
    \city{Sydney}
  \country{Australia}
}
\email{ruitao.feng@unsw.edu.au}
\author{Chengwei Liu}
\affiliation{%
  \institution{Nanyang Technological University}
  \city{Singapore}
  \country{Singapore}
}
\email{chengwei001@e.ntu.edu.sg}

\author{Kaixuan Li}
\affiliation{%
  \institution{East China Normal University}
  \city{Shanghai}
  \country{China}
}
\email{kaixuanli@stu.ecnu.edu.cn}
\author{Zhengzi Xu}
\affiliation{%
  \institution{Nanyang Technological University}
    \city{Singapore}
  \country{Singapore}
}
\email{zhengzi.xu@ntu.edu.sg}
\author{Liming Nie}
\affiliation{%
  \institution{Nanyang Technological University}
    \city{Singapore}
  \country{Singapore}
}
\email{liming.nie@ntu.edu.sg}
\author{Yang Liu}
\affiliation{%
  \institution{Nanyang Technological University}
    \city{Singapore}
  \country{Singapore}
}
\email{yangliu@ntu.edu.sg}

\author{Yixiang Chen}
\authornote{Corresponding author}
\affiliation{%
  \institution{East China Normal University}
  \city{Shanghai}
  \country{China}
}
\email{yxchen@sei.ecnu.edu.cn}

\begin{abstract}
  Quality assurance (QA) tools are receiving more and more attention and are widely used by developers. Given the wide range of solutions for QA technology, it is still a question of evaluating QA tools. Most existing research is limited in the following ways: \textit{(i)} They compare tools without considering scanning rules analysis. \textit{(ii)} They disagree on the effectiveness of tools due to the study methodology and benchmark dataset. \textit{(iii)} They do not separately analyze the role of the warnings. \textit{(iv)} There is no large-scale study on the analysis of time performance. To address these problems, in the paper, we systematically select 6 free or open-source tools for a comprehensive study from a list of 148 existing Java QA tools. To carry out a comprehensive study and evaluate tools in multi-level dimensions, we first mapped the scanning rules to the CWE and analyze the coverage and granularity of the scanning rules. Then we conducted an experiment on 5 benchmarks, including 1,425 bugs, to investigate the effectiveness of these tools. Furthermore, we took substantial effort to investigate the effectiveness of warnings by comparing the real labeled bugs with the warnings and investigating their role in bug detection. Finally, we assessed these tools' time performance on 1,049 projects. The useful findings based on our comprehensive study can help developers improve their tools and provide users with suggestions for selecting QA tools.
\end{abstract}










\maketitle
\section{Introduction}
Software quality has always been a continuing goal of developers and users. Due to the growing size and complexity of software, developers are facing a particularly difficult situation compared to that of the past few years. An increasing number of software bugs have been discovered during operations, resulting in severe consequences, including massive economic loss and even endangering human lives~\cite{aad,zh_2009, Lions1996ARIANE5F,secf}. Given the importance of software code quality, quality assurance (QA) tools have been widely used due to their low cost, convenience, and ability to find bugs~\cite{Ayewah_2007,nachtigall_large-scale_2022}.
 
Unfortunately, as a result of limited practical study, it is still challenging for users to evaluate and select the appropriate QA tools.
\textit{(1)} Recent research only compares the tools based on the recall or false positive of detection results and ignores one of the most important parts: \textit{Scanning rules}. Developers and users cannot understand the coverage and granularity of the scanning rules of the tools according to the existing studies~\cite{Sonarqube,SonarQubees,habib_how_2018,thung_what_2012,thung_what_2015,bu_evaluating_2018,zheng_value_2006,khendek_comparing_2005}. 
\textit{(2)} Evaluations of the tools based on the detection results are influenced by the benchmark dataset and methodology. Some research is conflicting when investigating tools' capabilities for identifying real bugs. Habib et al.~\cite{habib_how_2018} highlighted the poor recall of these tools on the Defects4J dataset~\cite{defects4j}. They found that a large majority (95.5\%) of the studied bugs were not detected. In contrast, Thung et al.~\cite{thung_what_2012,thung_what_2015} suggested that most bugs in the program can be reported by combining the results of these tools on different datasets and methods. 
\textit{(3)} Although there are some studies on the false positive warnings of QA tools, they focus on the truthfulness of warnings and how to reduce false positives~\cite{bu_evaluating_2018,zheng_value_2006,khendek_comparing_2005}. They pay no attention to the effectiveness of the warnings or whether they can provide clues for bugs. 
\textit{(4)} There is no large-scale study to analyze the time performance of the QA tools. 

\begin{figure}
\centering
    {\includegraphics[width=\columnwidth]{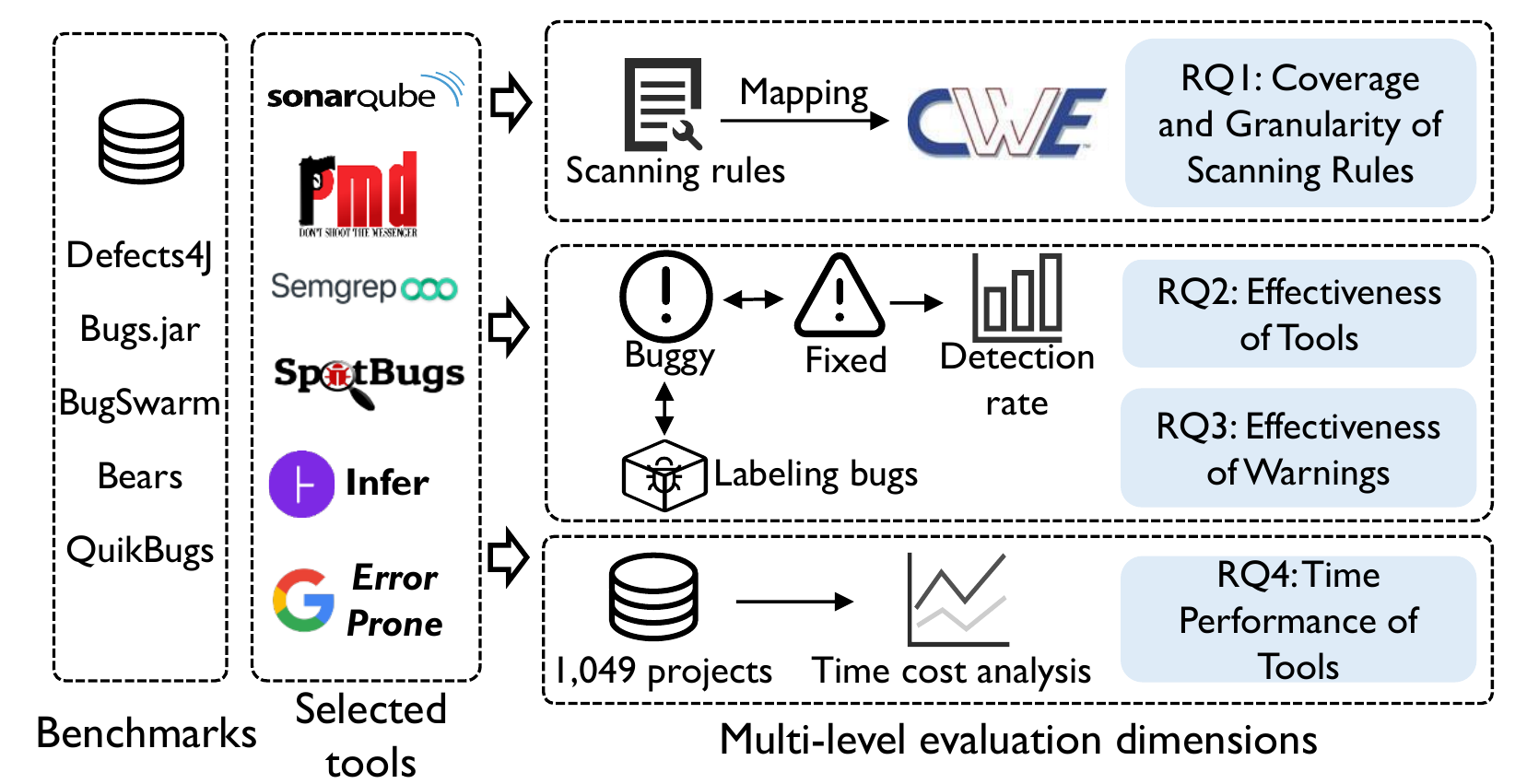}}
    \caption{Overview of our study}
    \label{overview}
\end{figure}

To address these problems, we conducted a comprehensive study on Java QA tools to compare and evaluate them on multi-level dimensions. The overview of our study is shown in Figure~\ref{overview}.
First, we selected 6 tools and 5 benchmarks for our study. 
\revised{Then, we analyzed the coverage and granularity of the rules of these tools through manual comparison. However, even when performing manual analysis, it is challenging to compare scanning rules directly because of the vast differences in how the rules are presented among these selected tools. We need to build a unified ground reference to make a fair comparison of the rules of these QA tools. Common Weakness Enumeration (CWE)~\cite{cwe} has been widely used to manage bugs in various software projects~\cite{AnalyzingJSS} and its weaknesses have been mapped to scanning rules of some tools (e.g., SonarQube~\cite{Sonarqube}, Semgrep~\cite{semgrep}). Moreover, ISO/IEC 5055~\cite{ISO_5055}, an international standard for measuring software quality, also uses CWE as a reference.}
Therefore, we used CWE as a reference and spent 6 person-month to establish a connection and exploring the gap between the rules and CWE to study the coverage and granularity between QA tools. After that, we conducted a benchmark experiment on 5 benchmarks to investigate the effectiveness of these tools. \revised{Note that, in this work, we focused on assessing bug detection ability, not bad coding practice and other issue detection ability of QA tools.} We compared the warnings reported by the tools in the buggy version and the fixed version to analyze the detection rate of the tools. Notably, we took substantial effort (1.5 person-month) to manually label the bug information, mapping it to CWE. On top of that, we studied the effectiveness of the warnings by comparing the labeling bug with the warnings and exploring their role in finding bugs. Finally, we evaluated the time performance of the tools in large-scale projects to analyze their time cost. 

Through our study, \textit{(1)} we find that the rules of SonarQube and Error Prone have a higher coverage than others. The granularity of rules in Infer, SpotBugs, and Semgrep is finer than others. \textit{(2)} For the detection rates of bugs, the tools with higher coverage of rules are proven that they have a higher detection rate, except for PMD. The detection rates range from 0.3\% to 10.2\%. The tools with finer granularity appear to be more focused on the bugs in one domain area. \textit{(3)} With the analysis of the warnings and labeled dataset, we find that most of the warnings are not the true reasons for bugs, especially the warnings of PMD. However, the warnings would be useful in finding bugs. \textit{(4)} Based on the analysis, Error Prone and SonarQube have the best capability in bug detection and Infer presents an excellent capability in detecting its focused issues on our benchmarks. \textit{(5)} Finally, we observe each tool requires an average of 65 seconds to execute a project analysis. The execution time of each tool varies greatly. In large projects, the execution times of SonarQube and Infer increase rapidly. Furthermore, the execution time of Semgrep is only moderately affected by project size. It is worth noting that although Error Prone runs with compilation, it is more efficient than 4 tools due to its method of embedding the compilation into checking. 

In summary, the contributions of this work are as follows:
\begin{itemize}

\item We conducted a benchmark experiment of 6 QA tools on 1,425 bugs from 5 benchmarks and a large-scale experiment of these tools on 1,049 projects. Our comprehensive study is the largest study on QA tools ever (i.e., $6 \times (1,425 \times 2 + 1,049 ) = 23,394$\footnote{1,425 bugs have both buggy version and fixed version.} scanning tasks). We spent over 4 months preparing and executing these projects. 

\item We spent 7.5 person-month mapping 1,813 scanning rules of 6 tools to CWE and 311 detected bugs by these tools to CWE. To the best of our knowledge, this is the first work that constructs a connection between references, rules, tools, and datasets.

\item We evaluated the selected QA tools from multi-level dimensions, including coverage and granularity of scanning rules, the effectiveness of tools, the effectiveness of warnings, and time performance. Our evaluation implements a systematic and comprehensive comparison of QA tools from scanning foundation to scanning results and scanning expense. The study data are released on the website: \url{https://sites.google.com/view/quality-tools-analysis/home}.

\end{itemize}

\section{Overview}
In this section, we first 
detail the criteria for tool selection and the used datasets in our study. We then introduce the method of mapping scanning rules to CWEs for each tool.

\subsection{Tool Selection}
In this study, we first collected 148 QA tools supporting Java language from 6 references~\cite{githubsat,owasp1,owasp2,Nist,wiki,Kompar} including GitHub, OWASP, NIST, Kompar, and Wikipedia, \revised{which is collected by snowballing the tool set from ~\cite{nachtigall_large-scale_2022}}. Then, we investigated the characteristics of the tools and defined 6 selection criteria as follows. \revised{For each criterion, our answer is either true or false.}
 
\begin{itemize}[leftmargin=*]
\item\textit{\textbf{Criterion \#1 (Available):}} The tool should be open-source or has a free release if it is a commercial tool.
\item\textit{\textbf{Criterion \#2 (Being maintained):}} The tool should still be maintained by the developers during the last two years so that we can exclude some outdated and unpopular tools. 

\item\textit{\textbf{Criterion \#3 (User-friendly command line interface):}} We need to run the tool for benchmarks and large-scale experiments, it needs to have a command line interface and execute without any limitations(e.g., can only run as a plugin).

\item\textit{\textbf{Criterion \#4 (Well documented with rules):}} As introduced, establishing the linkage between CWEs and the rules for detection is our first task. The tool should have its own integral public rules and be described well in the natural language. Otherwise, it is difficult to conduct the analysis based on scanning rules. 

\item\textit{\textbf{Criterion \#5 (Quality related):}} The tool should be relevant to code quality analysis/assurance such as bug detection. \revised{The tools should claim that they can detect ``\textit{quality issues}'', ``\textit{bugs}'', ``\textit{defects}'', ``\textit{flaws}'', ``\textit{mistakes}'', ``\textit{failure}'', or ``\textit{fault}'' in their documents.}

\item\textit{\textbf{Criterion \#6 (Not similar to other tools):}} The selected tool should be representative and partially unique, which means it should not be an integrated or evolved version of other similar tools in our collection.
\end{itemize}

Finally, followed by the defined criteria, we selected 6 QA tools for Java from 148 tools: SonarQube, SpotBugs, Error Prone, Infer, PMD, and Semgrep. Table~\ref{tool_profile} shows a brief introduction to these tools. We briefly describe each tool as follows.

\begin{itemize}[leftmargin=*]
\item\textit{\textbf{SonarQube}}~\cite{Sonarqube}. SonarQube is a code quality and security analysis tool by SonarSource. It is a multi-language supported tool, which consists of two parts: SonarScanner and SonarQube servers. In our study, we use version 9.5.0 of the SonarQube server and version 4.1.0 of SonarScanner.

\item\textit{\textbf{SpotBugs}}~\cite{Spotbugs}. This QA tool is licensed under the LGPL-2.1 license. It is the successor of Findbugs~\cite{findbugs} and got 2.8k stars and 482 forks on GitHub. SpotBugs aims to find bugs in a Java program. We use version 4.7.0.

\item\textit{\textbf{Error Prone}}~\cite{Errorprone}. Error Prone is also an open-source QA tool on Java developed by Google, which got 6k stars and 685 forks on GitHub. It is licensed under the Apache-2.0 license. Error Prone aims to catch common programming mistakes at compilation time. In this study, since most projects in benchmark require Java 8 compilation environment, we use version 2.10.0 of Error Prone, the latest version supporting Java 8 compilation.

\item\textit{\textbf{Infer}}~\cite{infer}. Facebook developed this QA tool and is now an open-source tool licensed under the MIT license. It got 4k stars and 1.3k forks on GitHub. Infer supports Java and C/C++/Objective-C language and aims to produce a list of potential bugs. We use version 1.0.0 of Infer.

\item\textit{\textbf{PMD}}~\cite{PMD}. PMD is an open-source source code analysis tool licensed under PMD's BSD-style license and gets 4k stars and 1.3k forks on GitHub. It claims that it can find common programming flaws in multi-language programs. In our study, we use version 6.41.0 of PMD.

\item\textit{\textbf{Semgrep}}~\cite{semgrep}. Semgrep is a code analysis tool supporting 25+ languages. It claims that it can find bugs and run security scans in programs. We use the open-source version 0.81.0 of Semgrep.


\end{itemize}
\begin{table}\small
\caption{Tool profile}
\begin{center}
\scalebox{0.88}{
\begin{tabular}{llccc}
\toprule
\textbf{Tools}                & \textbf{Aim}               & \textbf{Version} & \textbf{\# License}& \textbf{\# Rules} \\ \midrule
\textit{\textbf{SonarQube}}   & code quality analysis      & 9.5.0            & LGPL-3.0 & 552                       \\
\textit{\textbf{SpotBugs}}    & find bugs                  & 4.7.0            & LGPL-2.1 & 453                       \\
\textit{\textbf{PMD}}         & find flaws                 & 6.41.0           & PMD's BSD-style & 119                       \\
\textit{\textbf{Error Prone}} & catch mistakes             & 2.1.0            & Apache-2.0 & 405                       \\
\textit{\textbf{Infer}}       & find potential bugs        & 1.0.0            & MIT & 120                       \\
\textit{\textbf{Semgrep}}     & find bugs                  & 0.81.0           & LGPL-2.1 & 164                       \\ \bottomrule
\end{tabular}}
\end{center}
\label{tool_profile}
\end{table}

\subsection{Datasets}
We introduce two types of datasets, one for the benchmark-based experiment, and the other for the large-scale experiment.

\subsubsection{Benchmark Dataset\label{defects4jsection}}

For our study, we need high-quality benchmark datasets to evaluate the tools. To this end, we defined the following three criteria:
\begin{itemize}[leftmargin=*]
\item\textit{\textbf{Criterion \#1 (Java language):}} The benchmark should contain bugs in Java language projects.
\item \textit{\textbf{Criterion \#2 (Peer-reviewed):}} The benchmark should be peer-reviewed. It needs to be present in at least one research paper. 

\item\textit{\textbf{Criterion \#3 (Manually-fixed version):}} For each bug, the benchmark should have a manually-fixed version accordingly.
\end{itemize}

Specifically, we first collected 7 datasets by the criterion \#1 and \#2, including Bugs.jar~\cite{bugsdotjar}, iBUGS~\cite{ibugs}, Bears~\cite{bears}, IntroClassJava~\cite{IntroClassJava}, BugSwarm~\cite{bugswarm}, QuixBugs~\cite{quikbugs}, and Defects4J~\cite{defects4j}. Then we selected the datasets by criterion \#3. Finally, there are 5 datasets that meet all requirements: Bugs.jar, Bears, BugSwarm, QuixBugs, and Defects4J. 

Since the tools such as Infer, SpotBugs, SonarQube, and Error Prone, have a strong requirement for compilation, we requested that the projects in these datasets can be compiled. We excluded some non-compilable projects from the dataset. The final number of projects and bugs contained in each dataset is shown in Table~\ref{dataset_details}.


\begin{table}\small
\caption{Details of datasets}
\begin{center}
\begin{tabular}{lcc}
\toprule
\textbf{Dataset}    & \textbf{\# Project} & \textbf{\# Bugs} \\ \midrule
Defects4J  & 17        & 835    \\
Bugs.jar   & 6         & 371    \\
BugSwarm   & 35        & 108    \\
Bears      & 40        & 71     \\
QuixBugs   & 40        & 40     \\ \midrule
Total bugs &           & 1,425   \\ \bottomrule
\end{tabular}
\label{dataset_details}
\end{center}
\end{table}
\subsubsection{Large-scale Experiment Dataset} \revised{To assess the time performance of QA tools, we require a large-scale experiment in addition to a benchmark-based experiment. Our goal is to obtain the efficiency of executing these tools in real-world scenarios. Consequently, we need to obtain numerous projects of diverse sizes and developers, ensuring that the selection process is free from any perceived bias. In addition, we need to ensure that the dataset in the project is representative which means not a project which is meaningless and unused. Initially, as the compilation necessities of the tools allow projects released in package managers to be compiled more easily, we chose open-source projects from repositories in package managers such as Maven and Gradle. Subsequently, we applied two filters to the projects: the packages associated with the open-source projects have been depended on by packages from other providers and have new packages relying on them within the past three years. In this step, we chose around 10,000 projects. Lastly, given the performance and compilation requirements, we picked 1,049 compilable projects. The versions of these projects are the most recent versions accessible at that time.}


\subsection{Mapping Scanning Rules to CWEs}
To study the effectiveness of scanning rules, we investigate the coverage and granularity of the rules. However, due to the different forms and presentations, it is difficult to directly compare the coverage and granularity of the rules between different tools. We need to establish links and explore gaps between tools and a baseline and then answer the research questions according to the baseline. 

\subsubsection{Common Weakness Enumeration (CWE)}
CWE~\cite{cwe} is a community developed list of software and hardware weakness types. It is language-independent and uses a generic approach to describe the weaknesses of software and hardware. CWE has evolved over the years to its current location. Recently it has been expanded to include weaknesses from quality characteristics beyond security~\cite{ISO_5055}. Now it has 927 weaknesses and contains a variety of different views, including Software Development, Hardware Design, and Research Concepts. In order to study the detection capability of tools in different bug categories, we manually mapped the rules of different tools to CWEs. 

\begin{figure}
\centerline{{\includegraphics[width=0.45\textwidth]{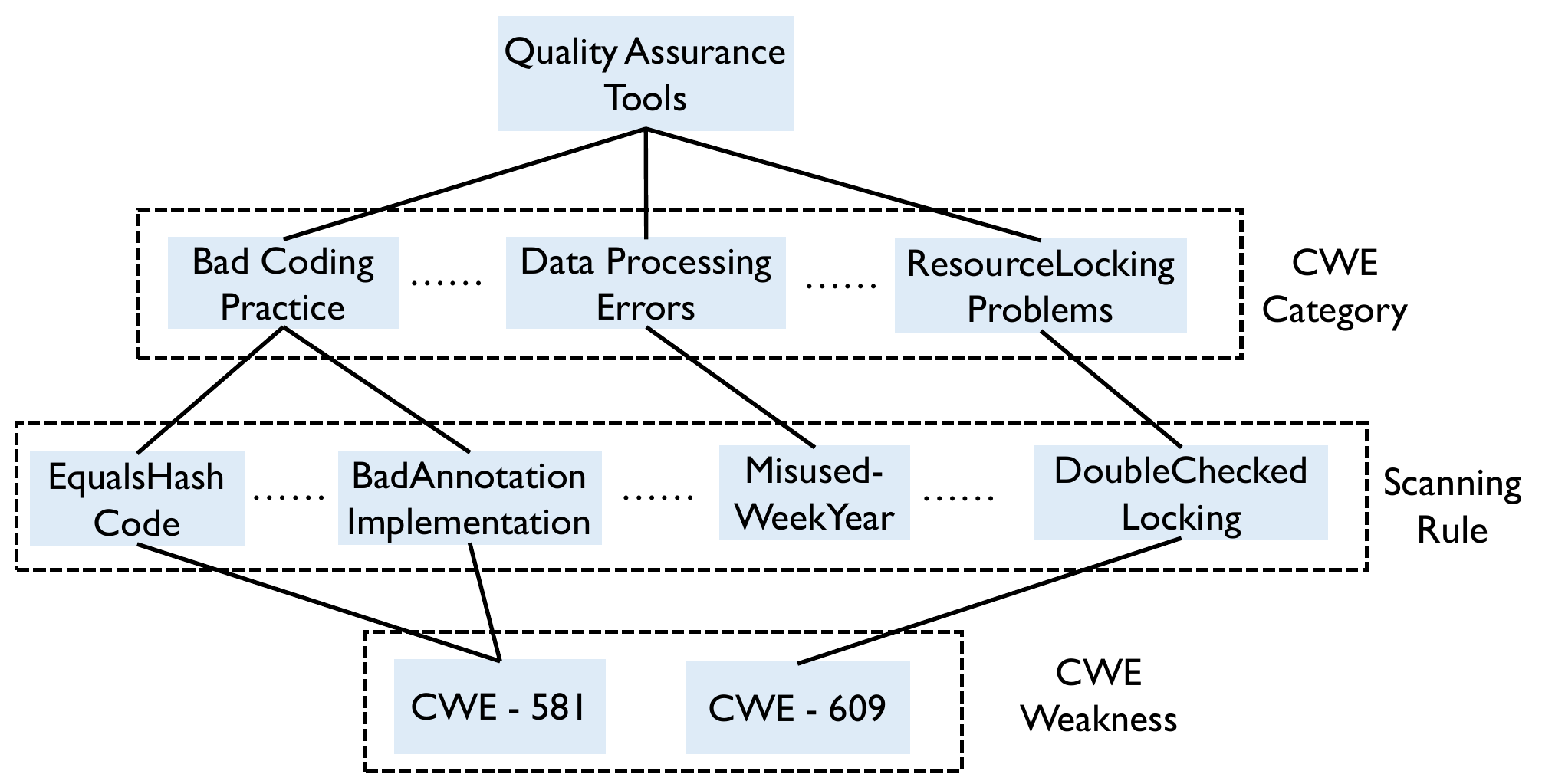}}}
\caption{The mapping structure of CWEs}
\label{cwe_fig}
\end{figure}

\subsubsection{Mapping Method}
In this study, we need to choose a CWE VIEW to map, which is a subset of CWE entries that provides a way of examining CWE contents. There are three main views in CWE, i.e. Software Development, Hardware Design, and Research Concepts. Since we focus on software code quality, we use the CWE VIEW: Software Development (ID:699)~\cite{cwesfd} instead of CWE VIEW: Research Concept (ID:1000)~\cite{cwerc} and CWE VIEW: Hardware Design (ID:1194)~\cite{cwehd} to map the rules. Moreover, Research Concept concentrates on the research of weaknesses. It aims to identify theoretical gaps within CWE systematically. It pays less attention to how to detect bugs, which is contrary to our research. Thus, CWE VIEW: Software Development is a better choice. This view contains the weaknesses and concepts for all aspects of software development at 418 weaknesses in total, including 40 CWE categories that contain a set of other CWE entries that share common characteristics.

The mapping process is structured into two parts. As shown in Figure~\ref{cwe_fig}, 
we need to establish the connections between scanning rules and CWE categories and the connections between scanning rules and CWE Weaknesses. (1) We mapped each scanning rule to the CWE category because the CWE categories contain a wide range of weaknesses. For example, for the rules of Error Prone \textit{``EqualsHashCode''} and \textit{``BadAnnotationImplementation''}, we mapped them to the CWE category \textit{``Bad Coding Practices''}. For another rule of Error Prone \textit{``MisusedWeekYear"}, we mapped it to CWE category \textit{``Data Processing Errors''}. (2) We tried to map the scanning rule to CWE weakness. Since the CWE types are more specific, there may be cases where the rules cannot be mapped to the CWE weakness. Moreover, there may also be more than one rule corresponding to a CWE weakness, as many scanning rules are finer than CWE weaknesses. For example, for the rules of Error Prone \textit{``EqualsHashCode''} and \textit{``BadAnnotationImplementation''}, we mapped them to CWE weakness \textit{``CWE-581: Object Model Violation''} belonging to the CWE category \textit{``Bad Coding Practices ''}. As for another rule \textit{``MisusedWeekYear''}, we failed to map it to CWE weaknesses belonging to the CWE category \textit{``Data Processing Errors''}. 

It is undeniable that there is subjectivity in these two parts. Furthermore, the mapping results are limited by the knowledge and experience of the mappers. To reduce the influence of subjective factors and the limitations of the knowledge and experience of the mappers, we used one-person mapping and two-person confirmation for this process. The mapping results will be accepted when the three authors' results are consistent. 

For each tool, we obtained the rules in default configuration from their official documentation or website, and the number of rules for each tool is shown in Table~\ref{tool_profile}. As for the SonarQube and Semgrep, they officially map the rules to CWE weaknesses. We confirmed their mapping result and map their rules to CWE categories. Finally, for each tool, we obtained a hierarchy similar to the one in Figure~\ref{cwe_fig}.




\section{Empirical Study}

\subsection{Research Questions}
In this study, we aim to answer the following research questions:

\subsubsection{\textbf{Coverage and Granularity of Scanning Rules (RQ1)}}
To what extent do scanning rules cover different bugs, and how do the granularity of scanning rules vary by tools?

For this RQ, we focus on the coverage and granularity of the scanning rules to investigate the effectiveness of these rules. We try to investigate the coverage of scanning rules for different bugs and the granularity gap of scanning rules in different tools.

\subsubsection{\textbf{Effectiveness of Tools (RQ2)}} To what extent can QA tools detect bugs from a diversity of benchmarks? 

For this RQ, we focus on the tools' detection rates for detecting bugs in a diversity of benchmarks to explore the capability of QA tools. 
Furthermore, we study the reasons for missing bugs to present improvement methods for QA tools. 

\subsubsection{\textbf{Effectiveness of Warnings (RQ3)}} How effective are the warnings reported by the QA tools? 

This RQ tries to investigate the gap between the warnings and the real source of bugs to study the effectiveness of the warning. In addition, we explore whether they can provide some hints for developers even if they are not the real source of bugs.

\subsubsection{\textbf{Time Performance of Tools (RQ4)}} What is the time performance of the QA tools? 

In this RQ, we focus on time performance to analyze the time cost of these tools. We compare the average time performance of each QA tool and their trend on different sizes of projects to explore the time cost of tool usage and the correlation between the technologies of tools and their execution time.

\subsection{Coverage and Granularity of Rules (RQ1)\label{RQ1}}
In this RQ, we analyze the coverage of scanning rules in various bug categories and the granularity of these rules in different tools to investigate whether they are well-designed. To this end, we mapped all rules to the CWEs. Table~\ref{rule_map} shows the number of mapped CWEs associated with the rules. More specifically, we summarize the top 5 CWE categories and concrete weaknesses from the results of analysis and present the results in Table~\ref{top_rule_cwe}. \revised{For each CWE category and CWE weakness, the numbers in parentheses indicate the quantity mapped to their rules.}
\begin{table}\small
\begin{center}
\caption{Rules mapping results}
\begin{tabular}{lcc}
\toprule
\textbf{Tools}                & \multicolumn{1}{c}{\textbf{\begin{tabular}[c]{@{}c@{}}\# mapped \\ CWE categories\end{tabular}}} & \multicolumn{1}{c}{\textbf{\begin{tabular}[c]{@{}c@{}}\# mapped \\ CWE weaknesses\end{tabular}}} \\ \midrule
\textit{\textbf{SonarQube}}   & 28                                                    & 41                                                    \\
\textit{\textbf{SpotBugs}}    & 22                                                    & 33                                                    \\
\textit{\textbf{PMD}}         & 19                                                    & 12                                                    \\
\textit{\textbf{Error Prone}} & 24                                                    & 21                                                    \\
\textit{\textbf{Infer}}       & 19                                                    & 28                                                    \\
\textit{\textbf{Semgrep}}     & 18                                                    & 22                                                    \\ \bottomrule
\end{tabular}
\label{rule_map}
\end{center}
\end{table}
\subsubsection{\textbf{Coverage of Rules}\label{coverrules}}
As shown in Table~\ref{rule_map}, SonarQube's rules cover the largest number of CWE categories, with 28 categories. The second is Error Prone, whose rules cover a total of 24 CWE categories. The third is SpotBugs, with 22 CWE categories. PMD, Infer, and Semgrep cover the 19, 19, and 18 CWE categories, respectively. By looking into the covered CWE categories (in Table~\ref{top_rule_cwe}), almost half of the rules in SonarQube, SpotBugs, PMD, and Error Prone belong to ``\textit{Bad Coding Practices}''. Especially for SonarQube, ``Bad Coding Practices'' accounts for 52.2\% of all rules. For Infer, as it claims that, it focuses on the null pointer dereferences, memory leaks, coding conventions, and unavailable APIs, 23\% of its rules focus on the ``\textit{Pointer Issues}'', 19.2\% concern ``\textit{Resource Management Errors}'', and 9.2\% focus on ``\textit{Memory Buffer Errors}''. As for Semgrep, over 26\% of the rules belong to ``\textit{Data Neutralization Issues}''.\\
\revised{\textit{\textbf{Missing Covered Categories.}}}
However, although all tools cover more than 18 CWE categories, there are still some categories that have not been included, such as \textit{``User Session Errors'', ``User Interface Security Issues'', ``Signal Errors'', ``Authorization Errors'', ``Communication Channel Errors'', ``Data Validation Issues'', and ``Credentials Management Errors''}. By exploring the tool implementations and the rule composition forms, we find the main reason for these uncovered categories is that some of them are designed for domain-specific issues. However, most of the scanning rules of tools are used to check for general errors. For domain-specific issues, such as web and user interface, developers need to configure and re-write the rules to adapt the tools to a specific domain. \\%
\revised{In addition, for a majority of tools, it is observed that most of the rules fall under a limited number of categories. Certain categories may contain a multitude of potential bugs, but these tools only implement a part of them. Therefore, some tools seem to cover a wide range of categories that, in reality, do not contain a substantial number of rule entries within each category.} For the 6 selected tools, the number of categories with less than 1\% of rules are: SonarQube with 17 categories, SpotBugs with 8 categories, PMD with 10 categories, Error Prone with 10 categories, Infer with 7 categories, and Semgrep with 5 categories. \\
\revised{\textit{\textbf{Comparative Coverage Focus of Different Tools.}}}
The coverage varies between different tools. For example, ``\textit{Pointer Issues}'' does not appear in the rules of PMD and Semgrep. ``\textit{Concurrency Issues}'' does not appear in the rules of Semgrep. On the contrary, ``\textit{Pointer Issues}'' is dominant in Infer's rules, and ``\textit{Concurrency Issues}'' plays an important role in SpotBugs. We further investigate the reasons behind this difference and find that:
\revised{Tools such as SonarQube, SpotBugs, and Error Prone apply a multitude of analysis methods. They not only scan the source code but also examine binary code or collect compile-time data, enabling them to gather more information and enhance their analytical capabilities. As a result, they implement more rules to detect bugs. In contrast, other tools like PMD and Semgrep can only construct abstract syntax trees, data flows, and control flows via source code analysis. This method provides substantially less information compared to the others, making the implementation of numerous rules challenging. Nevertheless, the objective of these tools is to swiftly identify issues on a large scale.}
Hence, they can only solve the issues under CWE categories which can be detected more easily, e.g., ``\textit{Bad Coding Practices}'', ``\textit{Complexity Issues}'', and ``\textit{Information Management Errors}''. As for the tool, Infer, its underlying formalism - Separation Logic is a mathematical method that can facilitate the reasoning about the program to find potential mutations in computer memory~\cite{SL}. Thus, the issues caused by null pointer dereference and leakage of resources and memory definitely attract more attention, which further leads to the dominance of the corresponding rules.

\smallskip

\noindent\fbox{
	\parbox{0.95\linewidth}{
		\textbf{Answer to RQ1:} \textit{The scanning rule coverage of QA tools needs to be improved. Even though SonarQube and Error Prone have a higher coverage (28 and 24 mapped CWE categories), they are more willing to detect easier categories, e.g., Bad Coding Practices. In addition, each tool has its own specific focus point. Users can combine the features of the tools with their own needs in practice {(e.g., select Infer for point issue detection)}.} 
	}
}

\subsubsection{\textbf{Granularity of Rules}}

According to the result of CWE weakness mapping (in Table~\ref{rule_map}), SonarQube still holds the lead with a total of 43 CWE weaknesses involved. Next are SpotBugs and Infer which involve 33 and 28 CWE weaknesses, respectively. Finally, Semgrep, Error Prone, and PMD involve 22, 21 and 12 CWE weaknesses, respectively. From this perspective, it is clear that both SonarQube and SpotBugs have the advantage of a higher coverage of the CWE weaknesses. For the purpose of studying the granularity of the rules, which is another focus point, we need to investigate the number of rules mapped to each CWE.

As shown in Table~\ref{top_rule_cwe}, we can observe that one CWE weakness can be mapped to many rules. For instance, CWE-476 can be mapped to 22 rules in our study. As we described in the $\S$~\ref{coverrules}, Infer focuses on checking issues caused by null pointer dereference, memory leakage, coding convention, and unavailable API. The root cause of CWE-476 happens to be null pointer dereference. As a result, Infer takes full advantage of its own strengths and achieves the best detection effectiveness on CWE-476. In fact, after a more concrete analysis of the used fundamental components in the tool, we find that most of these rules are detected by the same Infer detector, namely \textit{Pulse}. 
Pulse divides the null pointer dereference in a more fine-grained way, making its checks in this area more detailed. 
In addition, other CWE weaknesses (e.g., CWE-124, CWE-825) that can be mapped to Infer's rules are also related to similar risks caused by pointers, buffers, and resources. 

For Semgrep, CWE-611 and CWE-319 have the largest number of mapped rules, which is 16. Different from Infer, Semgrep does not use a specific technical implementation to detect a certain category of issues. 
Semgrep's rules development is jointly maintained by the open-source community, thereby, developers may want to subdivide issues (e.g., CWE-611 and CWE-319). Hence, Semgrep's rules are more fine-grained and detailed from the aspect of software security.

For SpotBugs, CWE-476 (null pointer dereference) has been mapped to 9 rules, which also shows a more fine-grained detection than other tools (e.g., SonarQube). SpotBugs divides the issues caused by null pointer dereference into 9 types, including null pointer dereference, null pointer dereference in method on exception path, etc. Similar to Infer, this tool also involves multiple detectors, most of which are mapped to a CWE weakness belonging to the same detector. For instance, issues associated to CWE-476 are actually detected by the detector, namely \textit{FindNullDeref}. After detection, SpotBugs finally breaks them down at the rule level to give the developer a more detailed description.

For the other 3 tools, SonarQube, Error Prone, and PMD, there is little difference in granularity at the mapped CWE weakness level. Their rules have the appropriate granularity relative to CWE weakness. {Therefore, the scanning rules of Infer, Semgrep, and SpotBugs have finer granularity than the rules of SonarQube, Error Prone, and PMD from CWE. On the one hand, users who want to find detailed reasons (e.g., junior programmers) for the bugs should use tools with finer granularity (i.e., Infer, Spotbugs, Semgrep). On the other hand, users who prefer a comprehensive bug detection without caring about the details (e.g., skillful programmers) can use tools with coarser granularity (i.e., SonarQube, Error Prone, PMD).}
\revised{However, due to the possibility that the rules of the tools may be too finer or too coarse, the rules can not be mapped to CWEs, and there is still a significant gap between them and CWE. This also indicates that there are missing rules in these tools, causing them to not fully cover CWE. Therefore, scanning rules can still be improved.}

\begin{table}
\begin{center}
\caption{Top 5 CWE categories and CWE weaknesses in rules}
\scalebox{0.75}{
\begin{tabular}{lll}
\toprule
\textbf{}                                & \multicolumn{1}{l}{\textbf{CWE category}}                                               & \textbf{\begin{tabular}[c]{@{}c@{}}CWE\\ weakness\end{tabular} } \\ \midrule
\multirow{5}{*}{\rotatebox{90}{\textit{\textbf{SonarQube}}}}  & Bad Coding Practices (288, 52.2\%)                                                               & CWE-476 (5)           \\
                                              & Error Conditions, Return Values, Status Codes (43, 7.8\%)                                       & CWE-546 (2)           \\
                                              & Expression Issues (36, 6.5\%)                                                                   & CWE-396 (2)           \\
                                              & Permission Issues (28, 5.1\%)                                                                   & CWE-477 (2)           \\
                                              & Data Processing Errors (25, 4.5\%)                                                              & CWE-595 (2)           \\ \midrule
\multirow{5}{*}{\rotatebox{90}{\textit{\textbf{SpotBugs}}}}   & Bad Coding Practices (187, 41.3\%)                                                               & CWE-476 (9)           \\
                                              & Concurrency Issues (42, 9.3\%)                                                                  & CWE-125 (5)           \\
                                              & Data Processing Errors (41, 9.1\%)                                                              & CWE-908 (5)           \\
                                              & Permission Issues (38, 8.4\%)                                                                   & CWE-1024 (4)          \\
                                              & API/Function Errors (20, 4.4\%)                                                               & CWE-248 (4)           \\ \midrule
\multirow{5}{*}{\rotatebox{90}{\textit{\textbf{PMD}}}}        & Bad Coding Practices (44, 37.0\%)                                                               & CWE-252 (2)           \\
                                              & Complexity Issues (18, 15.1\%)                                                                   & CWE-609 (1)           \\
                                              & Error Conditions, Return Values, Status Codes (13,10.9\%)                                       & CWE-1339 (1)          \\
                                              & Permission Issues (7, 5.9\%)                                                                   & CWE-1051 (1)          \\
                                              & Data Processing Errors (6, 5.0\%)                                                              & CWE-570 (1)           \\ \midrule
\multirow{5}{*}{\rotatebox{90}{\textit{\textbf{Error Prone}}}}& Bad Coding Practices (155, 38.3\%)                                                               & CWE-570 (4)           \\
                                              & Data Processing Errors (73, 18.0\%)                                                              & CWE-1024 (4)          \\
                                              & Error Conditions, Return Values, Status Codes (31, 7.7\%)                                       & CWE-595 (3)           \\
                                              & API/Function Errors (21, 5.2\%)                                                               & CWE-805 (2)           \\
                                              & String Errors (15, 3.7\%)                                                                       & CWE-581 (2)           \\ \midrule
\multirow{5}{*}{\rotatebox{90}{\textit{\textbf{Infer}}}}      & Pointer Issues (28, 23.3\%)                                                                      & CWE-476 (22)           \\
                                              & Resource Management Errors (23, 19.2\%)                                                          & CWE-124 (7)           \\
                                              & Complexity Issues (12, 10.0\%)                                                                   & CWE-502 (6)           \\
                                              & Resource Locking Problems (11, 9.2\%)                                                           & CWE-825 (4)           \\
                                              & Memory Buffer Errors (11, 9.2\%)                                                                & CWE-413 (4)          \\ \midrule
\multirow{5}{*}{\rotatebox{90}{\textit{\textbf{Semgrep}}}}    & Data Neutralization Issues (42, 25.6\%)                                                          & CWE-611 (16)           \\
                                              & Data Processing Errors (25, 15.2\%)                                                              & CWE-319 (16)          \\
                                              & Cryptographic Issues (23, 14.0\%)                                                                & CWE-502 (8)           \\
                                              & Information Management Errors (18, 11.0\%)                                                       & CWE-89 (7)           \\
                                              & Resource Management Errors (16, 9.8\%)                                                          & CWE-94 (6)           \\ \bottomrule
\end{tabular}}
\label{top_rule_cwe}
\end{center}
\end{table}


\noindent\fbox{
	\parbox{0.95\linewidth}{
		\textbf{Answer to RQ1:} \textit{From the mapping results, the granularity of Infer, Semgrep, and SpotBugs' rules is finer than that of SonarQube, Error Prone, and PMD. This is achieved by detailing the result of their sub-detectors. But the gap between tools and CWE is still quite wide since only a small part of the rules are successfully mapped to CWE weaknesses.} 
	}
}

\subsection{Effectiveness of Tools (RQ2)\label{RQ2}}
We investigate the coverage and granularity of rules for each tool in $\S$ \ref{RQ1}.
In this RQ, we investigate the effectiveness of QA tools in detecting real-world bugs. 
To achieve this, we conducted bug detection using 6 tools on 5 benchmarks, Defects4J, Bugs.jar, Bears, BugSwarm, and QuixBugs.


\subsubsection{Experiment Setup}

In the experiment, we first checked out all buggy versions and fixed versions in benchmarks. 
For the 6 tools, each tool has its specific requirements for execution. 
Error Prone and Infer have to conduct bug checking during compilation. 
SpotBugs and SonarQube need the binary files generated after compilation.
PMD and Semgrep need to check the source files. 
To fairly test the tools, we compiled all projects and collected all their dependencies and binary class file directories. We input the tools with compilation, binary files, and source files, respectively.
\revised{In the benchmark-based experiments, we tried to discern bug-sensitive warnings. These are warnings that appear in the buggy versions but not in the corresponding fixed versions. We analyzed both the projects in the buggy and fixed versions. Then we generated a report for each tool per project for subsequent evaluation. 
If a tool reports a bug in the buggy version and does not report the same bug following the fix, we categorize this as a bug-sensitive warning and conclude that it accurately predicts the presence of the bug. We believe that using bug-sensitive warnings to verify the correction of the warning is more effective than simply comparing the bug lines. This method not only contains the line information in it but also identifies all the related warnings of the bugs.}


\subsubsection{Result and Analysis of Effectiveness}
To understand the size of bugs in the benchmarks, we count the number of files affected and the number of line changes between buggy and fixed versions. 
\revised{The findings reveal that 84.7$\%$ of bugs only required alterations to a single file, and 72.3$\%$ of bugs necessitated changes to fewer than ten lines of code. This suggests that the majority of bugs in a project are typically confined to a single file and can be rectified with a few line modifications.}



The bug detection results of each tool are presented in Table~\ref{result_num}. We observe that the result vary widely across different tools. In general, the detection rates of each tool are not satisfied. Even for the tool with the best results, PMD only detects 10.2\% (146/1,425) bugs. SonarQube has a similar detection rate with PMD, which detects 9.5\% of bugs. The next is Error Prone and SpotBugs, with 7.6\% and 6.2\%, respectively. Infer and Semgrep have lower detection rates, which only detect 1.3\% and 0.3\% of 1,425 bugs, respectively.


\begin{table}\footnotesize
\caption{Tool detection results}
\begin{center}
\scalebox{0.84}{
\begin{tabular}{lcccccc}
\toprule
\multirow{2}{*}{\textbf{Tools}}  & \multicolumn{5}{c}{\textbf{\# Bugs Detected in Different Benchmarks}}                            \\ \cline{2-7} 
                                 & \textbf{Defects4J} & \textbf{Bugs.jar} & \textbf{BugSwarm}& \textbf{Bears}  & \textbf{QuixBugs} &\textbf{Total}\\ \midrule
\textit{\textbf{SonarQube}}      & \cellcolor{gray!130}\color{white}68 (8.1\%)          & \cellcolor{gray!180}\color{white}40 (10.8\%)    &\cellcolor{gray!130}\color{white} 21 (19.4\%)     & \cellcolor{gray!180}\color{white}3 (4.2\%)              &\cellcolor{gray!180}\color{white} 4 (10.0\%)    &\cellcolor{gray!130}\color{white} 136 (9.5\%)     \\
\textit{\textbf{SpotBugs}}       &\cellcolor{gray!50} 44 (5.3\%)          & \cellcolor{gray!50}18 (4.9\%)    &\cellcolor{gray!130}\color{white} 21 (19.4\%)     & 0 (0.0\%)               & \cellcolor{gray!130}\color{white}3 (7.5\%)  & \cellcolor{gray!50}86 (6.0\%)         \\
\textit{\textbf{PMD}}            & \cellcolor{gray!180}\color{white}90 (10.8\%)         & \cellcolor{gray!80}30 (8.1\%)     & \cellcolor{gray!180}\color{white}22 (20.4\%)      & \cellcolor{gray!180}\color{white}3 (4.2\%)             & \cellcolor{gray!80}1 (2.5\%)     &\cellcolor{gray!180}\color{white} 146 (10.2\%)     \\
\textit{\textbf{Error Prone}}    & \cellcolor{gray!80}61 (7.3\%)          & \cellcolor{gray!180}\color{white}40 (10.8\%)   & \cellcolor{gray!50}5 (4.6\%)       & \cellcolor{gray!80}2 (2.8\%)               & \cellcolor{gray!80} 1 (2.5\%)     &\cellcolor{gray!80} 109 (7.6\%)     \\
\textit{\textbf{Infer}}          &\cellcolor{gray!20} 10 (1.2\%)          & \cellcolor{gray!20}4 (1.1\%)     &\cellcolor{gray!20} 4 (3.7\%)       & \cellcolor{gray!50} 1 (1.4\%)               & 0 (0.0\%)         &\cellcolor{gray!20} 19 (1.3\%) \\
\textit{\textbf{Semgrep}}        & 0 (0.0\%)           & 0 (0.0\%)      &\cellcolor{gray!20} 4 (3.7\%)    & 0 (0.0\%)                 & 0 (0.0\%)       & 4 (0.3\%)   \\ \midrule
\textbf{Total Bugs} & 835                & 371           & 108         & 71                       & 40        & 1,425        \\ \bottomrule
\end{tabular}}
\label{result_num}
\end{center}
\end{table}

\begin{figure}
\centerline{\includegraphics[width=0.90\columnwidth]{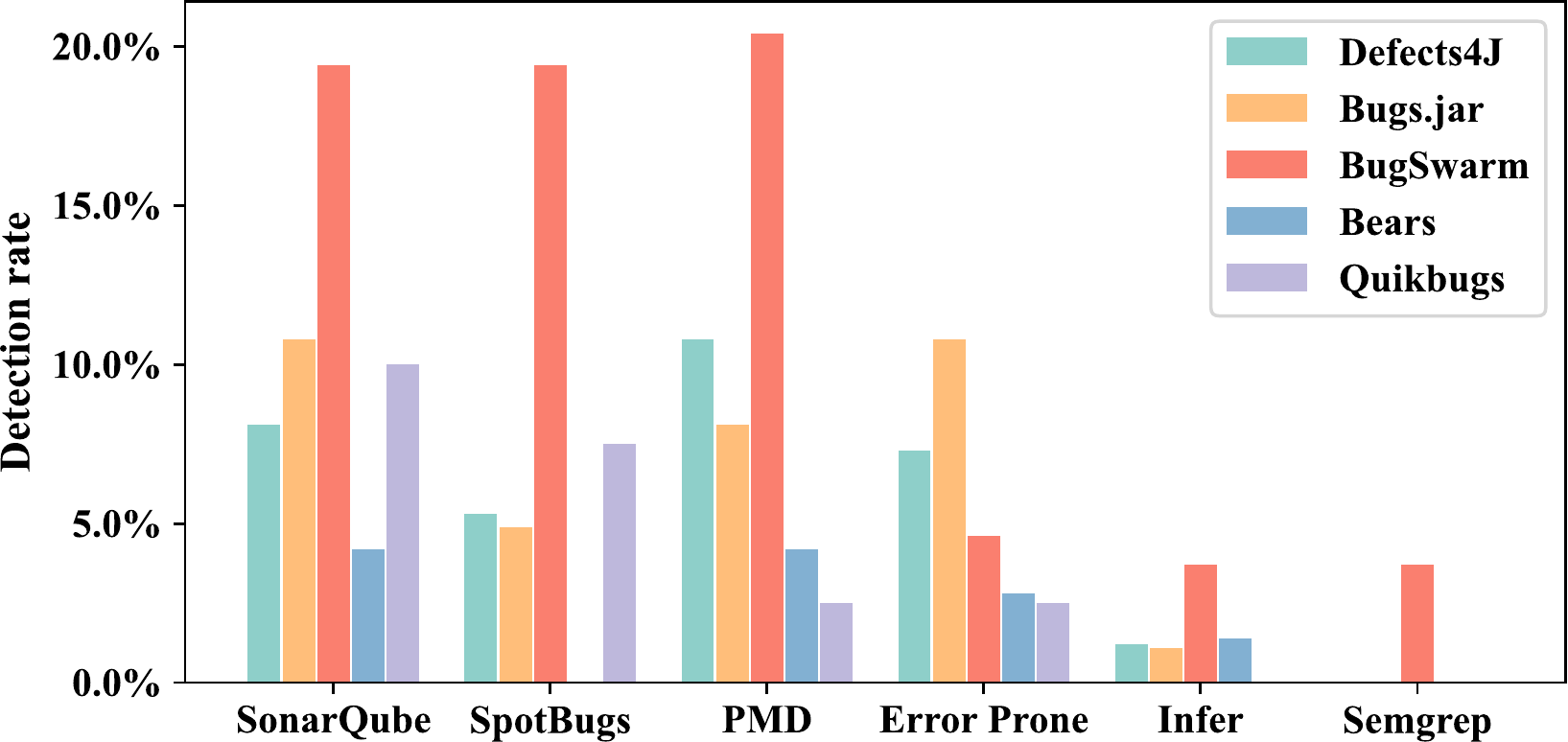}}
\caption{Detection results in different benchmarks}
\label{resultbenchmark}
\end{figure}

The effectiveness of each tool differs across different benchmarks. As shown in Figure~\ref{resultbenchmark}, SonarQube detects much more bugs (19.4\%) in BugSwarm than other benchmarks while it detects the least (4.2\%) in Bears. Regarding other benchmarks, SonarQube detects around 10\% of bugs. As for SpotBugs, it detects most (19.4\%) in BugSwarm while it can not detect bugs in Bears. For other benchmarks, it detects around 5.5\% of bugs. PMD also shows a better detection capability (20.4\%) in BugSwarm. In other benchmarks, it presents an unstable detection rate from 2.5\% to 10.8\%. Error Prone detects most in Bugs.jar, which is different from other tools, and its detection rate in other benchmarks ranges from 2.5\% to 7.3\%. As for Infer and Semgrep, they detect a little in the benchmarks. Noted that, BugSwarm is still detected most in Infer and Semgrep.
The main reason that leads to the difference is that different benchmarks have different categories of bugs. As shown in $\S$~\ref{RQ1}, the coverage of the scanning rules varies between tools, thus leading to the difference in detection results.
\revised{Overall, although the bug pattern may cover the influence scope of bugs in benchmarks, the detection rates of the tools are disappointing based on their claims and their rule coverage.} 
Among 6 tools, PMD and SonarQube demonstrate relatively better capability in bug detection. Error Prone and SpotBugs also can detect some bugs, but they are less capable than PMD and SonarQube. Since Infer and Semgrep focus on point issues and security issues, they do not perform well on the benchmark we have collected. The results of the benchmark experiments agree most with the answers to RQ1. SonarQube, which has been confirmed with higher coverage in RQ1, shows better capabilities than other tools. Error Prone and SpotBugs are also more effective than Infer and Semgrep. However, despite the lower coverage of PMD, it detects the most among the 6 tools. The reason is that the warnings reported by PMD are not most useful or relevant. It achieves this effectiveness by reporting a large number of warnings. We will discuss this extensively in RQ3. 


\smallskip

\noindent\fbox{
	\parbox{0.95\linewidth}{
		\textbf{Answer to RQ2:} \textit{Our experiment on 5 benchmarks shows that the QA tools can not detect bugs as expected. The best tools, PMD and SonarQube, can detect only around 10\% of bugs. Next are Error Prone and SpotBugs, which can detect 7.5\% and 6\% of bugs, respectively. Semgrep and Infer, which focus more on a certain type, show a lower detection rate in our experiment.} 
	}
}


\subsubsection{Reasons for Missing Detection of Bugs}
To understand why more than 80\% of bugs are not detected by these tools, we manually inspect the code and reports of the missed bugs. We conducted a random sample analysis on 30 bugs that are missing detection by 6 tools and summarize three reasons for missing detection of bugs.


First, many bugs (19 of 30) contain specific domains, which cannot be detected by existing tools. For example, in the Time project in Defect4J with bug No.16, which is not detected by any tool, a variable ``iDefaultYear'' is used to overwrite the given instant year, which leads to the Default Year is not a leap year. Therefore, it cannot correctly parse the date 29, February. However, due to the specialty of this leap year issue, it is difficult to be detected by common rules.
\revised{Secondly, a certain number of missing bugs (7 out of 30) occur exclusively in exceptionally particular circumstances. These conditions are overlooked by both QA tools and developers. Nevertheless, these particular conditions may be the ones users employ in the real world, triggering the bugs. For example, in the project Commons-Math in Defects4J with bug No.1, here, the constructor Fraction will create a fraction by the given double value and maximum denominator. However, an overflow exception is thrown when a fraction is initialized from a double that is very close to a simple fraction. Despite many tools possessing rules for detecting overflow, this represents an incredibly specific situation where the gap between implementation and invocation is considerable. Consequently, it becomes challenging for tools to manage this scenario effectively.
Thirdly, a subset of these bugs (4 out of 30) embodies logical or algorithmic issues that evade identification by rule-based detection methods such as pattern matching or other forms of static analysis.    
For instance, bug No.21 in the Commons-Math project from Defects4J fails to accurately calculate the rank of the matrix under certain conditions. This error stems from an incorrect matrix transformation due to misarranged columns, representing a profound logical mistake. Such an error is challenging to encapsulate within a rule. The majority of QA tools are not designed or capable of detecting these types of logical errors.}

\smallskip

\noindent\fbox{
	\parbox{0.95\linewidth}{
		\textbf{Findings in RQ2:} \textit{\revised{The main reasons for the missing detection of bugs are insufficient scanning rule coverage, neglect of highly specific scenarios, and the inability to handle logical or algorithmic errors, which are not the target of most QA tools. These insights can be utilized to enhance the effectiveness of these tools in the future.}} 
	}
}

\subsection{Effectiveness of Warnings (RQ3)\label{RQ3}}
In RQ2, we investigate the effectiveness of tools by comparing the warnings between the buggy and fixed versions. We find PMD detects more bugs than other tools with fewer rules which is far away from our expectations. In this RQ, we take a further step into the gap between the warnings and the real bugs (ground truth) to study the practical precision of the reported warnings. In addition, we explore whether they can provide any hints for fixing when not able to give actual reasons for the bugs.

\subsubsection{Datasets Annotation}
In order to answer RQ3, we need to know the actual reasons for the bugs detected by the tools.
Therefore, we manually labeled our datasets by mapping bugs to CWEs. The mapping method is similar to that of mapping rules. Firstly, we reviewed the bug issues, bug report, and some other documents of the bugs, and compared the source code in the fixed and buggy versions. Next, we mapped them to CWE categories. If the corresponding CWE weakness is not identified, we only map this bug to the CWE category. Otherwise, the bug will be mapped to the CWE weakness. In this process, we followed the procedure that one person works on mapping, and two persons validate the results as a check. Finally, we summarize the top 5 CWE categories and weaknesses from the mapped results of detected projects and present them in Table~\ref{proj_cwe}.

\begin{table}\small
\caption{Top 5 CWE categories and weaknesses in detected projects}
\begin{center}
\scalebox{1}{
\begin{tabular}{ll}
\toprule
 \textbf{CWE category}               & \textbf{CWE weakness} \\ \midrule
 Data Processing Errors (54)           & CWE-1339 (6)             \\
 Numeric Errors (38)                   & CWE-1024 (3)             \\
 Behavioral Problems (22)              & CWE-476 (3)              \\
 Bad Coding Practices (18)              & CWE-353 (2)              \\
 Encapsulation Issues (16)               & CWE-835 (2)              \\ 
                                    \bottomrule
\end{tabular}}
\end{center}
\label{proj_cwe}
\end{table}

\subsubsection{Analysis of Warnings}
By comparing the CWE categories mapped to the bugs with those mapped to the warning rules, we find that only a small percentage of the CWE categories are actually detected. Through the experiments, we observe only Error Prone, SonarQube, SpotBugs, and PMD successfully detect bugs. Specifically, Error Prone and SonarQube detect the largest number of bugs at 12, followed by PMD at 8 as the second. SpotBugs detect 4 bugs, and Infer detect 3 bugs. In this way, the capability of PMD decreases significantly, which explains our doubts raised in the previous discussion: PMD does not have a large number of rules, yet it performs very well on most datasets. \revised{ Most of the warnings reported by PMD are not the actual reasons for specific bugs while it claimed that it can find flaws. Instead, most of them are just referred to as \textit{Bad Coding Practice} }. On one hand, some of PMD's rules do successfully detect bugs. On the other hand, although its warnings are bug-sensitive, they are not actually the root cause of the bugs. As Infer focuses on a specific domain, the results show a good ability in finding targeted issues due to its implementation. For SonarQube, Error Prone, and SpotBugs, their detection rate decreases sharply, but the relationship between their capabilities and rules is as expected. In detail, the ability of SpotBugs and the quality of its rules are not as good as Error Prone in our experiments.

We also count the number of warning rules reported by each tool, and the CWE categories and weaknesses in detected bugs to observe the most frequently reported warnings. Table~\ref{cwe_num} shows the top 5 reported rules and CWE categories and weaknesses, which the rules map to. Compared the numbers in Table~\ref{proj_cwe} with those in Table~\ref{cwe_num}, there is a wide gap between the warnings reported by tools and the actual reasons for bugs. SonarQube, SpotBugs, PMD, and Error Prone all report a large number of \textit{Bad Coding Practice} in the warnings. However, the fact is most bugs in the benchmark are \textit{Data Processing Errors}. PMD also reports a number of \textit{Complexity Issues} and \textit{Documentation Issues} which would not be the primary reason for the bug either. Because Semgrep focuses on security issues, it presents a worse capability of bug detection on our benchmark. 

\smallskip
\noindent\fbox{
	\parbox{0.95\linewidth}{
		\textbf{Answer to RQ3:} \textit{Compared with the real reasons for bugs, only a few warnings are effective, and most of them refer to Bad Coding Practice. Although some of the warnings are bug-sensitive, they are not actually the real cause of the bugs. Comparatively, Error Prone and SonarQube have the best capability in bug detection. Infer presents an excellent capability in detecting its focused issues.} 
	}
}
\smallskip
\begin{table*}
\caption{Top 5 rules, CWE categories, and CWE weaknesses in benchmark experiment}
\begin{center}
\scalebox{0.73}{
\begin{tabular}{llll}
\toprule
\multicolumn{1}{l}{\textbf{}}            & \multicolumn{1}{c}{\textbf{Rule}}          & \multicolumn{1}{c}{\textbf{CWE category}}        & \multicolumn{1}{l}{\textbf{CWE weakness}} \\ \midrule
\multirow{5}{*}{\rotatebox{90}{\textit{\textbf{SonarQube}}}}   & Deprecated code should be removed (25)       & Bad Coding Practices (190)                        & CWE-563 (29)                           \\
                                            &Local variables should not be declared and immediately returned or thrown (16)&Data Processing Errors (38)& CWE-595 (2)                  \\
                                            & \"@Deprecated\" code should not be used (15)    & API / Function Errors (13)                         & CWE-1024 (1)                            \\
                                            & Unused assignments should be removed (14)  & Documentation Issues (10)                                & CWE-1069 (1)                            \\
                                            & The diamond operator ("<>") should be used (14)  & Error Conditions, Return Values, Status Codes (4)  &                             \\ \midrule
\multirow{5}{*}{\rotatebox{90}{\textit{\textbf{SpotBugs}}}} & EI\_EXPOSE\_REP (14)                           & Bad Coding Practices (51)                     & CWE-476 (13)\\
                                            & EI\_EXPOSE\_REP2 (10)                            & Permission Issues (39)                        & CWE-821 (6)\\
                                            & DLS\_DEAD\_LOCAL\_STORE (10)                     & Error Conditions, Return Values, Status Codes (15)  & CWE-1188 (5)                            \\
                                              & THROWS\_METHOD\_THROWS\_CLAUSE\_BASIC\_EXCEPTION (8) & Pointer Issues (13)         & CWE-786 (2)                            \\
                                              & IS2\_INCONSISTENT\_SYNC (6)    & Initialization and Cleanup Errors (6) &  CWE-477 (15)                                   \\ \midrule
\multirow{5}{*}{\rotatebox{90}{\textit{\textbf{PMD}}}}         & UnnecessaryImport (166)                       & Complexity Issues (228)                         & CWE-563 (14)                           \\
                                              & UselessParentheses (29)                         & Bad Coding Practices (103)                          & CWE-476 (6)                                      \\
                                              & ControlStatementBraces (29)                      & Documentation Issues (16)                         & CWE-546 (5)    \\
                                              & GuardLogStatement (21)                           & Error Conditions, Return Values, Status Codes (14) & CWE-459 (5)            \\
                                              & UnnecessaryLocalBeforeReturn (17)                & Memory Buffer Errors (6)                        & CWE-563 (29)                                      \\ \midrule
\multirow{5}{*}{\rotatebox{90}{\textit{\textbf{Error Prone}}}}& MissingOverride (47)                             & Bad Coding Practices (117)                       & CWE-595 (2)\\
                                            & UnusedVariable (29)                                & Data Processing Errors (38)                        & CWE-1024 (1)\\
                                                & InvalidParam (18)                             & API / Function Errors (13)                         & CWE-1069 (1)                           \\
                                              & UnnecessaryParentheses (11)                      & Documentation Issues (12)                        &                           \\
                                              & DefaultCharset (9)                          & Error Conditions, Return Values, Status Codes (4) &                                       \\ \midrule
\multirow{4}{*}{\rotatebox{90}{\textit{\textbf{Infer}}}}       & NULL\_DEREFERENCE (13)                           & Pointer Issues (13)                                & CWE-476 (13)                            \\
                                              & THREAD\_SAFETY\_VIOLATION (10)                      & Concurrency Issues (10)                            &                        \\
                                                & RESOURCE\_LEAK (4)                                & Resource Management Errors (4)                  &                              \\
                                                & INEFFICIENT\_KEYSET\_ITERATOR (2)               & Complexity Issues (2)                             &                                   
                                            \\ \midrule
\multirow{5}{*}{\rotatebox{90}{\textit{\textbf{Semgrep}}}}      & hardcoded\_api\_key (8) &  Key Management Errors (8)   & CWE-532 (3)\\
                                                & android\_logging (3)& Information Management Errors (3) & CWE-611 (2)\\
                                                & documentbuilderfactory\-disallow\-doctype\-decl\-missing (1)& Data Processing Errors (2)& CWE-94 (1)\\
                                                & imports.owasp.java.xxe.possible.import.statements (1)& Data Neutralization Issues (1)\\
                                                & script-engine-injection.script-engine-injection (1)&  \\ \bottomrule
\end{tabular}}
\label{cwe_num}
\end{center}
\end{table*}

Although most reported warnings are not real bugs, we are interested in their actual effectiveness as they are bug-sensitive. We suppose that these warnings may be caused by the complications of the bugs and they can provide some hints for developers as guidance to understand the bugs.

In this part, we also manually inspect the warnings associated with the source code of bugs. We inspect a random sample of 50 warnings in the detected projects and find there are three types of warnings. (1) The warnings (37 of 50) are a suggestion for improving maintainability or performance. It provides a few clues to bugs. The fixed version removes this part of the code as a coincidence. These warnings include useless parentheses, unused import, missing override etc. For example, in bug "WICKET-5569\_5efb8091" of benchmark Bug.jar, PMD reported a warning about useless parentheses. In the fixed version of the bug, the warning disappeared due to the completely refactored code. (2) The warnings (11 of 50) belonging to \textit{Bad Coding Practice} can really provide hints for finding bugs, especially for those variants that are not using private functions. For example, in the project JacksonDatabind with bug No.29 of benchmark Defects4J, it is caused by the next token of variable \textit{p2} may be a null value. There is a warning reported by PMD that the variable \textit{t} has never been used and the next token of variable \textit{p2} is variable \textit{t}. The warning does not directly point out the root reason for the bugs, but we can notice that the next token of variable \textit{p2} may be null according to the warnings. In this regard, PMD performs better, but developers may need some experience to find bugs with the help of the warning information. (3) The warnings (2 of 50) are indeed potential danger. These warnings may include null point dereference. For example, regarding project Jsoup with bug No.4 of benchmark Defects4J, Infer reported a warning that \textit{m.group(1)} may be null, which may not be the reason for this bug. However, \textit{m.group(1)} is assigned to String \textit{name} which will be a parameter called by \textit{full.containsKey()}. If \textit{name} is null, \textit{containsKey} will throw an \textit{NullPointerException}.

\smallskip
\noindent\fbox{
	\parbox{0.95\linewidth}{
		\textbf{Findings in RQ3:} \revised{\textit{While some warnings reported by the tool are not indicative of actual bugs, a portion (26\%) of these warnings (e.g., null point dereference) do provide valuable insights for identifying bugs within certain contexts. Nevertheless, a significant percentage of these warnings (74\%), such as useless parentheses, prove to be of limited usefulness in pinpointing specific bugs.}}
	}
}

\subsection{Time Performance (RQ4)}
\subsubsection{Experiment Setup}

To fairly compare the performance of these tools, we first cloned the 1,049 well-selected projects as the experiment objects, and based on them, we execute each tool independently to capture their execution time on each project.
Next, considering that these tools have different requirements when examining user projects, i.e., different requirements on inputs, we prepared the execution environment and measured their execution times separately for each tool. 
Specifically, (1) for tools (SpotBugs and SonarQube) that take binaries (i.e., jar files and class files) as inputs (SonarQube requires both binaries and source code), all these projects have to be built and compiled before experiments. However, since such a process could be heavily influenced by the time of downloading dependencies and network traffic, we only measure the execution time after the projects are compiled. (2) For tools (PMD and Semgrep) that simply scan source code, we directly feed them with the source code of the 1,049 projects. (3) For tools that intercept the compilation processes to identify quality issues, we are unable to split the execution times of Infer and Error Prone out of the compilation time, and we can only record the time of the entire compilation process. Instead, we prepare the binary files and dependencies before compiling the projects with javac command for Infer and Error Prone to minimize the influence of compilation.

Our experiments are conducted on a server with 80 vCPUs (Intel(R) Xeon(R) Gold 6248 CPU @ 2.50GHz $\times2$) and 188G of RAMs. Furthermore, 
we also copy the projects into the RAM disk to fasten the experiment process 
and prevent the influence of 
cloudy disk reading/writing. The speed of the RAM is 2,933 MT/s.


\begin{table}
\caption{Execution time for each tool}
\begin{center}
\scalebox{0.79}{
\begin{tabular}{llllc}
\toprule
\multirow{2}{*}{\#} & \multicolumn{1}{l}{\multirow{2}{*}{\textbf{Tools}}} &\multicolumn{1}{l}{\multirow{2}{*}{\textbf{\textbf{Execution requirement}}}} & \multicolumn{2}{c}{\textbf{Execution time}}   \\ 
                    & \multicolumn{1}{c}{}                     &  & \textbf{Average}  & \multicolumn{1}{c}{\textbf{Total}} \\ \midrule
1                   & \textit{\textbf{PMD}} & Source code                                       & 00:00:07 & 02:13:59 \mybar{7.7}       \\
2                   & \textit{\textbf{Error Prone}} & Execution with compilation                              & 00:00:18 & 05:24:14 \mybar{18.4}      \\
3                   & \textit{\textbf{SpotBugs}}  & Binary file                                 & 00:00:18 & 05:24:43 \mybar{18.6}      \\
4                   & \textit{\textbf{SonarQube}} & Source code and binary file                                  & 00:00:24 & 07:16:04 \mybar{24.9}      \\
5                   & \textit{\textbf{Infer}}      & Execution with compilation                                 & 00:00:55 & 16:03:19 \mybar{55.1}       \\
6                   & \textit{\textbf{Semgrep}}   & Source code                                 & 00:04:25 & 77:26:01 \mybar{265.7}       \\ \midrule
\multicolumn{2}{c}{\textbf{Average per tool}}                                       & & 00:01:05 & \multicolumn{1}{l}{18:57:23}           \\ \bottomrule
\end{tabular}}
\end{center}
\label{exec_time}
\end{table}

\subsubsection{Results and Time Performance Analysis}

Table \ref{exec_time} shows the average and total execution time of each tool on 1,049 selected projects. 
Generally, it takes 65 seconds averagely for each tool to scan a project.
PMD is the fastest tool which only averagely takes 7 seconds to scan a project. Apart from PMD, Error Prone, SpotBugs and SonarQube can finish their scanning within 30 seconds per project, while Infer and Semgrep are the two that take the longest time to execute (i.e., 55 seconds and over 4 minutes, respectively).


Besides, We find tools that intercept compilation processes to retrieve intermediate information are not as expected to be slower than tools that simply take binary files as inputs (i.e., not integrated with compilation). For instance, Error Prone identifies bugs during compilation, but its average execution time is less than the SpotBugs and SonarQube which simply take the binary files as inputs. 
However, Infer, another tool that is required to be executed along with compilation, doubles the execution time of SpotBugs and SonarQube.
Such a difference in execution time between Error Prone and Infer is probably because that
Error Prone captures bugs by checking the intermediate products along with compilation, while Infer only retrieves related information into its own intermediate language and reasons potential bugs with more complex techniques (i.e., analysis based on abstract interpretation) after compilation. This also explains why Infer achieves better results in detecting bugs that are more complicated (i.e., null pointer dereference, memory leakage, unavailable API, etc.) in previous experiments. \revised{Naturally, some analyses performed by Infer, such as Pulse analysis, are not as resource-intensive if employed iteratively - that is, analyzing commits where the preceding commit has already undergone analysis.}
\smallskip
\noindent\fbox{
	\parbox{0.95\linewidth}{
		\textbf{Answer to RQ4:} \textit{On average, each tool takes 65s to scan a project, while the execution time of each tool varies (7s to 265s). Error Prone which intercepts compilation processes to scan is not as expected to be slower than most of the tools that simply take binary files as inputs, while Infer is much slower due to its complex analysis.} 
	}
}
\smallskip

Furthermore, we notice that tools (PMD and Semgrep) that detect bugs by directly scanning source code have completely different performances. According to Table \ref{exec_time}, PMD is the fastest tool, while Semgrep is the slowest one. To further demystify the reason for the difference, we further classify the projects into different clusters by LOCs (Lines of code) of Java files.

\begin{figure}
\centerline{\includegraphics[width=0.85\columnwidth]{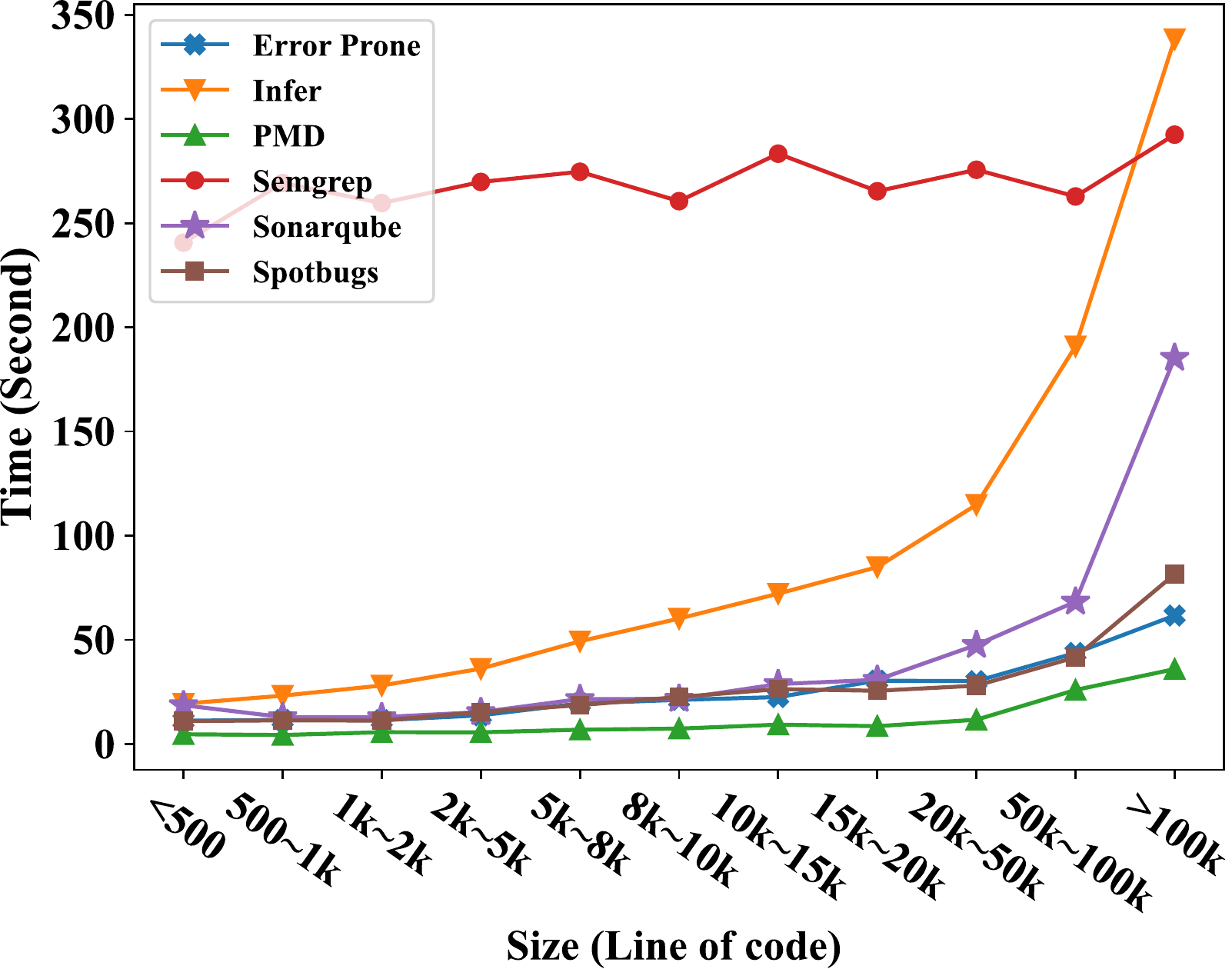}}
\caption{Tool execution time with project size}
\label{run_time}
\end{figure}

Figure~\ref{run_time} shows the average execution time of each tool on the project with different LOCs. 
Generally, the execution times of most of these tools are heavily influenced by the size of projects except Semgrep. PMD is always the fastest one regardless of the LOCs of projects.
SpotBugs, Error Prone, and SonarQube have similar performance when the projects are small (i.e., less than 50K LOCs), but the execution time of SonarQube rises rapidly when the projects are larger. 
Besides, the execution time of Infer is mostly influenced by the size of examined projects, and it is the slowest tool on projects with over 100K LOCs.
For Semgrep, the results show that the project size can barely influence its execution time, and it takes around 265 seconds to examine projects regardless of their size. 

This is because Semgrep splits a project into small components and parallels the scanning tasks on different components to reduce the influence of large projects, which makes it insensitive to the size of projects. However, such an approach sacrifices some of the accuracy (i.e., some inter-procedure communications could be lost during the scans), leading to false negatives of bugs. Moreover, unlike other QA tools, in order to have flexible and addable rules, Semgrep fails to execute the rules in parallel and needs to initialize each rule first, which makes the execution time much longer than others even if on a small project. For instance, although the execution time for each rule of Semgrep is pretty fast, it still takes 265 seconds on average to examine each project against 164 rules in our experiment. Such findings indicate the potential direction of high parallel on not only the execution of rules but also the detection of sliced projects when designing new QA tools in the future.

\smallskip
\noindent\fbox{
	\parbox{0.95\linewidth}{
		\textbf{Findings in RQ4:} \textit{The tool execution is influenced by the size of examined projects in different degrees, PMD is always the fastest tool, while Semgrep is barely influenced by the size of projects due to its unique paralleled processing of splitted project components.} 
	}
}

\section{Discussion}
\subsection{Lessons Learned}
We discuss actionable suggestions for the different stakeholders to improve, select, and use QA tools.

\noindent\textbf{For Developers of QA Tools.} Existing QA tools cannot effectively detect bugs as expected, we conclude with some tips for developers to improve the tools: (1) Increase the coverage of scanning rules, especially for the domain-specific rules. (2) Provide users with a flexible and simple method to adjust the granularity of rules and augment the rules. (3) Combine the different tools which have a specific focus (e.g., Infer) to increase the detection rate and reduce false positives. (4) Optimize the imbalance caused by the scanning rule in parallelism and scanning module in parallelism and automatically adjust the method according to the different sizes of projects.

\noindent\textbf{For Users of QA Tools.} To improve software quality, we suggest users select and use QA tools according to the following guidelines: (1) In specific domains, users should not rely on the original scanning rule. On the one hand, they can select tools that augment rules (e.g., PMD and Semgrep) and rewrite domain-specific rules to execute QA tools. On the other hand, they can choose tools with a specific focus. For example, they can select Infer to detect point issues. (2) In a security-related area, users should use the static application security testing tools to analyze vulnerability instead of general QA tools. (3) In other areas, users can select tools with higher coverage of rules (e.g., SonarQube). Noted that, when the projects are enormous, Sonarqube is not a good choice.

\subsection{Threats to Validity}
\noindent\textbf{External Validity.} A potential threat to external validity is related to the fact that the datasets we used as benchmark may not be an accurate representation of all bugs in software development. We tried to reduce this threat by collecting 5 different datasets and selecting constantly updated and widely-used datasets. Another concern is that although we tried to establish 6 criteria to select tools, the selection of the QA tools may not be representative and popular enough. Moreover, the projects we used in large-scale experiments may be biased and not large enough. We attempted to select more projects and reduce the filtering criteria to minimize threats. 
There is also an external threat from the analysis of the time performance of each tool. We attempted to exclude the interference of external factors in the time count, and we put the analysis items into the RAM disk and tried to reduce CPU usage at the same time.

\noindent\textbf{Internal Validity.} A potential threat to internal validity refers to the rule mapping to CWE manually. In this study, we mapped the rules to CWE categories and CWE weaknesses, which is a task needing expert knowledge and experience. Although we tried to reduce the subjectivity of mapping by one-person mapping and two-person confirmation, it could not eliminate the mapping errors and subjectivity. Apart from that, the process of manually labeling the benchmark suffers from the same threat with the scanning rules mapped. We also reduce this threat by cross-validation. The last threat is that the benchmark may contain bugs that have not been disclosed so far, and the selected tools can detect some new bugs which are still disclosed. However, we consider it feasible to draw valid findings by only focusing on known and existing bugs in the dataset, and discussing the warnings sensitive to the fixed versions.

\balance

\section{related work}
\subsection{Studies of Quality Assurance Tools}
Most current studies on QA tools only focused on the detection effectiveness of the tools. On the one hand, some works studied the precision and recall of tools. For example, Habib et al.~\cite{habib_how_2018} discussed the recall of 3 QA tools (SpotBugs, Error Prone, and Infer) based on the Defects4J dataset. They used a new method that manually validates each detected bug and found that the tools miss 95.5\% of bugs. Thung et al.~\cite{thung_what_2012}~\cite{thung_what_2015} also focused on the recall of 5 tools, FindBugs, JLint, PMD, CheckStyle, and JCSC, based on iBUGS Dataset. They automatically matched the warning with the buggy lines and found the tools could detect most bugs. Tomassi~\cite{tomassi_bugs_2018} inspected the capability of Error Prone and SpotBugs with a new benchmark dataset BugSwarm. It is a preliminary work with findings similar to~\cite{habib_how_2018}. Our work is different from theirs: (1) We collected 5 benchmarks to evaluate the recall of 6 selected QA tools which is more than twice the size of the previous largest dataset. Owing to the comprehensive set of benchmarks and tools, our study results are more representative in practice. (2) We apply a new method that compares the warning with the manual labeling dataset to investigate the real effectiveness of the warnings reported by QA tools. (3) Apart from the detection rate and the reasons for missing bugs, our findings in the benchmark experiment also present the specific focus of different QA tools.

On the other hand, some work presented the analysis of the warnings reported by QA tools. For example, Lu et al.~\cite{bu_evaluating_2018} analyzed the validity of 5 C/C++ warnings and concentrated on how machine learning methods can be used to improve the validity of warnings. Zheng et al.~\cite{zheng_value_2006} studied 3 C/C++ QA tools in three large industrial software systems by using an orthogonal defect classification scheme. They found that the number of the results of the tools can be effective in identifying problematic modules. Wagner et al.~\cite{khendek_comparing_2005} evaluated 3 Java QA tools (FindBugs, PMD, and QJ Pro) on several projects from the industry. They believed that different tools are complementary and could be used in an integrated way. Rutar et al.~\cite{rutar_comparison_2004} analyzed 5 QA tools(Bandera,
ESC/Java, FindBugs, JLint, and PMD) on Java projects. They found that there was no substitution between the different tools, and developed a meta-tool to combine their results. Compared with their studies, (1) our study focused on evaluating the effectiveness of warnings of Java QA tools by comparing the real reasons for bugs with the warnings in open-source benchmarks. (2) We conduct a manual analysis on investigating the help of the warnings in finding bugs. 

There are a few studies on the scanning rules. Lenarduzzi et al.~\cite{sqrules} investigated the fault-proneness of the SonarQube rules by comparing the classification results of machine learning models. They confirmed the SonarQube rules have a low ability for bug prediction. The most difference between theirs and ours is that we investigate the coverage and granularity of the scanning rules by
manual mapping CWE for 6 QA tools.


\subsection{Studies of Other Analysis Tools}
Some empirical studies concentrate on other analysis tools for detecting specific bugs~\cite{tomassi_real-world_2021,aloraini_evaluating_2017,fan2018large,chen2020empirical,chen2022ausera,shi2022large}. One of these studies considered null pointer exceptions~\cite{tomassi_real-world_2021}. Aloraini focused on the other specific bug: buffer errors. They analyzed the effectiveness of the static analysis tools on the buffer errors~\cite{aloraini_evaluating_2017}. Another section of the static analysis tools research focuses on security, vulnerabilities, and cryptography. Lipp et al. analyzed the effectiveness of vulnerability detection in C and C++ tools~\cite{lipp_empirical_2022}. Cheirdari et al. studied the false positive trends of static analysis tools on vulnerability detection~\cite{cheirdari_analyzing_2018}. Braga et al. analyzed the effectiveness of static analysis tools in identifying cryptography-related vulnerabilities~\cite{braga_practical_2017,braga_understanding_2019}. Furthermore, there exist many studies focusing on usability research regarding static analysis tools~\cite{nachtigall_large-scale_2022,vassallo_context_2018,johnson_study_2012,vassallo_how_2020,panichella_would_2015,do_why_2022} and analysis of static analysis tools for smart contracts~\cite{durieux_empirical_2020,ghaleb_how_2020}.
\section{conclusion}
In this paper, we presented a comprehensive study on 6 Java QA tools in multi-level dimensions. To better understand the coverage and granularity of the scanning rules of the tools, we mapped a total of 1,813 rules to CWE. Based on selected benchmarks, we conducted a benchmark experiment to reveal the effectiveness of tools. We also mapped 311 bugs to CWE to investigate the effectiveness. Finally, we conducted a large-scale experiment on 1,049 projects to analyze the time performance. Our study unveils many useful findings, including the comparison of the coverage and granularity between scanning rules of different tools, detection rate and reasons for missed bugs, the role of warnings in bug detection, execution time, and reasons for the difference between tools. We hope the findings are helpful and informative for developers and users.

\section*{Acknowledgments}
This research is supported by the East China Normal University Graduate Student International Conference Special Fund, Natural Science Foundation of China and the Israel Science Foundation (NSFC-ISF) Joint Program (62161146001, 3420/21), China Scholarship Council (202106140088,202206140052),  Cyber Security Cooperative Research Centre (CSCRC), Australia, National Research Foundation, Singapore, the Cyber Security Agency under its National Cybersecurity R\&D Programme (NCRP25-P04-TAICeN), the National Satellite of Excellence in Trustworthy Software Systems (NSOE-TSS) project under the National Cybersecurity R\&D (NCR) Grant award no. NRF2018NCR-NSOE003-0001, and the National Research Foundation Singapore and DSO National Laboratories under the AI Singapore Programme (AISG Award No: AISG2-RP-2020-019). Any opinions, findings and conclusions or recommendations expressed in this material are those of the author(s) and do not reflect the views of National Research Foundation, Singapore and Cyber Security Agency of Singapore.

\clearpage

\bibliographystyle{ACM-Reference-Format}
\bibliography{CameraReady}
\end{document}